\documentclass[prl,twocolumn,showpacs,superscriptaddress,
preprintnumbers,amsmath,amssymb,tightenlines]{revtex4}
\usepackage{graphicx}
\newcommand{\beq}{\begin{equation}}
\newcommand{\eeq}{\end{equation}}
\newcommand{\bea}{\begin{eqnarray}}
\newcommand{\eea}{\end{eqnarray}}
\newcommand{\ba}{\begin{array}}
\newcommand{\ea}{\end{array}}
\newcommand{\bc}{\begin{center}}
\newcommand{\ec}{\end{center}}

\newcommand{\ibid}{{\it ibid. }}
\newcommand{\bml}{\begin{mathletters}}
\newcommand{\eml}{\end{mathletters}}
\newcommand{\commentout}[1]{{}}
\newcommand{\bk}{{\bf k}}

\newcommand{\half}{\hbox{$1\over2$}}

\newcommand{\eq}[1]{(\ref{#1})}
\newcommand{\etal} {{\it et al.\/}}
\newcommand{\comment}[1]{{}}
\newcommand{\vol}[1]{{\bf #1}}

\begin{document}
\title{Simple mean-field theory for a zero-temperature Fermi gas at a
Feshbach resonance}
\author{Juha Javanainen}
\affiliation{Department of Physics, University of Connecticut,
Storrs, CT 06269-3046}
\author{Marijan Ko\v{s}trun}
\affiliation{Department of Physics, University of Connecticut,
Storrs, CT 06269-3046}
\author{Matt Mackie}
\affiliation{Department of Physics, Temple University, Philadelphia, PA 19122}
\author{Andrew Carmichael}
\affiliation{Department of Physics, University of Connecticut,
Storrs, CT 06269-3046}
\date{\today}

\begin{abstract}
We present a simple two-channel mean field theory for a zero-temperature
two-component Fermi gas in the neighborhood of a Feshbach resonance.
Our results agree with recent experiments on the bare-molecule
fraction as a function of magnetic field~[Partridge
\etal, cond-mat/0505353]. Even in this strongly-coupled gas of
${}^6$Li, the experimental results depend on the structure of the
molecules formed in the Feshbach resonance and, therefore, are not
universal.
\end{abstract}
\maketitle


{\em Introduction.}--Magnetoassociation creates a molecule from a pair of
colliding atoms when one of the spins flips in the presence of a magnetic
field tuned near a Feshbach resonance
\cite{STW76,TIE93}. Nevertheless, the ultracold community as a whole has
only reluctantly acknowledged the role of molecules in Feshbach-resonant
interactions, clinging instead to an all-atom single-channel model
solely characterized by a tunable $s$-wave scattering
length~\cite{KITP}. In this Letter we promote a simple
molecule-explicit two-channel mean-field theory to replace
all-atom models. We find an excellent agreement with an
experiment~\cite{PAR05} that should present a severe test for theories of
all ilk.

Our approach is built around two species of fermionic atoms that
magnetoassociate into bosonic ``bare" molecules. In the
experiments~\cite{PAR05}, a laser spectroscopy probe is applied that
detects bare molecules only. Below resonance, where
thermodynamics favors molecules, the challenge for single-channel
theories is to produce two types of molecules out of the same atoms, bare
molecules seen by the probe and something else. Instead, our two-channel
model simply states that bare molecules come ``dressed'' with atom
pairs.  Second, a typical all-atom theory rests on the notion of
universality, whereby the interactions between atoms can be lumped into a
scattering length and microscopic details of the physics do not matter.
Our approach also mimics universal behavior on the atom side of the
resonance for strong atom-molecule coupling. However,
the experiments~\cite{PAR05} cover a wide magnetic-field range on both
sides of the resonance and, according to our results, certain molecular
physics details are needed to satisfactorily match theory and
the experiments. Universality does not hold despite strong
interactions. Elementary as our mean-field theory is, it neatly bypasses
two limitations of all-atom theories.

{\em Model.}--As before~\cite{JAV04,MAC05}, we study two fermionic species
of atoms ($c_{{\bf k}\uparrow}$ and $c_{{\bf k}\downarrow}$) that
magnetoassociate into bosonic molecules ($b_{\bf k}$) within a
version of the Bose-Fermi model~\cite{RAN85} used to describe
superconductivity beyond the usual~\cite{EAG69} weak-coupling limit. The
corresponding microscopic Hamiltonian is
\bea
{H\over\hbar} &=& \sum_{\bf k}[(\half\epsilon_{\bf k}+\delta)
b^\dagger_{\bf k}b_{\bf k}+ \epsilon_{\bf k}(c^\dagger_{{\bf
k}\uparrow}c_{{\bf k}\uparrow}+c^\dagger_{{\bf k}\downarrow}c_{{\bf
k}\downarrow})]
\nonumber\\&&
+\sum_{{\bf k},{\bf k}'}\kappa_{\bk,\bk'}(b^\dagger_{{\bf k}+{\bf k}'}
c_{\bk\downarrow}c_{\bk'\uparrow} +
c^\dagger_{\bk'\uparrow}c^\dagger_{\bk\downarrow} b_{{\bf k}+{\bf k}'}
)\,.
\label{HAM}
\eea
Here $\hbar\epsilon_\bk=\hbar^2 k^2/2m$ is the kinetic energy for an atom
with wave vector \bk,
$\delta$ is the detuning controlled by the magnetic field ($\delta>0$
corresponds to an open dissociation channel for the molecules), and
$\kappa_{\bk,\bk'}$ are the atom-molecule coupling coefficients for two
atoms with momenta $\hbar\bk$ and $\hbar\bk'$.

The key approximation is to treat the boson
operators $b_{\bf k}$ and $b^\dagger_{\bf k}$ as classical conjugate
variables. Then the hierarchy of the equations of motion for fermion
correlation functions truncates exactly at the level of pair
correlations. As a technical approximation, we only keep the zero
momentum bosons, i.e., a uniform molecular condensate. We write fermion
occupation numbers and pair correlations as a function of free-atom energy
$\hbar\epsilon\equiv\hbar\epsilon_\bk$ as
$P(\epsilon)=\langle c^\dagger_{{\bf k}\downarrow}c_{{\bf
k}\downarrow}\rangle$=$\langle c^\dagger_{{\bf k}\uparrow}c_{{\bf
k}\uparrow}\rangle$, $C(\epsilon) =
\langle c_{-\bk\downarrow}c_{\bk\uparrow} \rangle$, and, given the
invariant atom number
$N$ arising from the conserved quantity $\hat{N}=\sum_{\bf k}(2
b^\dagger_{\bf k}b_{\bf k}+c^\dagger_{{\bf k}\uparrow}c_{{\bf
k}\uparrow}+c^\dagger_{{\bf k}\downarrow}c_{{\bf
k}\downarrow})$, define the molecular amplitude
$\beta=\sqrt{2/N}b_0$ such that $\beta^2$ is the fraction of atoms
converted into molecules. The result is the equations of
motion~\cite{JAV04}
\bea
i\dot{C}(\epsilon)&=&2\epsilon\,C(\epsilon)
+
    \hbox{$1\over\sqrt2$}{\Omega}\label{BEQEQ}\,
\beta f(\epsilon)[1-2P(\epsilon)]\,,\\
i\dot{P}(\epsilon)&=&
\hbox{$1\over\sqrt2$}\,\Omega f(\epsilon)[\beta
C^*(\epsilon)-\beta^*C(\epsilon)]\,,\\
i\dot\beta &=& \delta\beta+ \frac{3\,\Omega }
     {{2\sqrt{2}}\,{{{\epsilon }_F}}^{{3}/{2}}}
\int
d\epsilon\,\sqrt{\epsilon}\, f(\epsilon)C(\epsilon)\,.\label{INTEGRAL}
\label{FINEQ}
\eea
Here $\Omega=\sqrt{N}\kappa_{0,0}$ is the atom-molecule Rabi
frequency, and $f(\epsilon)$ [with
$f(0)=1]$ conveys the energy dependence of atom-molecule
coupling. Since the coupling coefficient
$\kappa_{0,0}$ is inversely proportional to the
square root of the quantization volume
$V$, the Rabi frequency is proportional to the square root of the
invariant density
$\rho=N/V$. Finally,
$\hbar\epsilon_F=\hbar^2(3\pi^2\rho)^{2/3}/2m$ is the usual Fermi
energy for a two-component gas which, so far, arises simply because the
sum over wave vectors is replaced by an integral over the frequencies.

The objective is to find the steady state of
Eqs.~\eq{BEQEQ}-\eq{INTEGRAL} at zero temperature. These equations admit
solutions of the form
$C(\epsilon;t)\equiv C(\epsilon)e^{-2i\mu t}$, $\beta(t) \equiv \beta
e^{-2i\mu t}$ and $P(\epsilon;t) \equiv P(\epsilon)$, where $\hbar\mu$
and $2\hbar\mu$ are recognized as the chemical potentials for atoms
and molecules. In fact, there are  too many solutions. We therefore
impose the condition of {\it maximal pairing}, whereby atoms only come
in pairs of
${\bf k}\!\!\uparrow$ and $-{\bf k}\!\!\downarrow$. This is technically
the usual BCS ansatz, but here it is not so much  an ansatz as a
natural consequence of how a pair of atoms couples to a molecule. In
mathematical terms we have
$|C(\epsilon)|^2=P(\epsilon)-P^2(\epsilon)$. Equations~\eq{BEQEQ}
and~\eq{INTEGRAL} then give the counterpart of the gap
equation in BCS theory,
\begin{equation}
\delta-2\mu={3\Omega^2\over8\epsilon_F^{3/2}}\int
d\epsilon\, f^2(\epsilon)\left(
{\sqrt{\epsilon}\over\sqrt{(\epsilon-\mu)^2+\Delta(\epsilon)^2}}
-{1\over\sqrt{\epsilon}}
\right)\,,
\label{GAPEQ}
\end{equation}
where $\Delta(\epsilon)=f(\epsilon)\beta\Omega/\sqrt{2}$ is the analog
of the pairing gap. As we~\cite{MAC02,JAV04,MAC05} and  others have
done in both boson~\cite{KOK02} and fermion~\cite{TIM01} systems, the term
$\propto 1/\sqrt{\epsilon}$ has been added into the integrand manually
and the corresponding (possibly infinite) constant on the left-hand
side has been absorbed into a ``renormalization'' of the
detuning $\delta$. One more equation is needed, and is found from the
conservation of atom number $\hat{N}$ that holds for the mean-field
equations~\eq{BEQEQ}-\eq{INTEGRAL} as well. We have
\begin{equation}
\beta^2+{3\over 4\epsilon_F^{3/2}}\int
d\epsilon\,\sqrt{\epsilon}\,\left(1-{\epsilon-\mu\over
\sqrt{(\epsilon-\mu)^2+\Delta(\epsilon)^2}}\right)=1\,.
\label{NORMEQ}
\end{equation}

\commentout
{
Our derivation is unusual in that we start from time dependence
and include the coupling profile $f(\epsilon)$, but overall,
Eqs.~\eq{GAPEQ} and~\eq{NORMEQ} to solve the quantities $\beta$ and $\mu$
are a garden variety two-channel mean-field theory for Boson-Fermion
coupling\cite{JAV04,MAC05,RAN85,TIM01}.
For the most part, we solve the theory numerically~\cite{JAV05}.
}

So far our approach deals with bare atoms and
molecules that would emerge if the Feshbach-resonance interaction were
suddenly turned off. In more practical terms, the sample breaks up
into bare atoms and molecules if the magnetic field is switched quickly
enough far enough away from the Feshbach resonance. The rate of change of
the detuning that separates quick from slow is
$|\dot\delta|\simeq\Omega^2$~\cite{MAC02,JAV04}, and fast switching may be
a tall order for the 834~G Feshbach resonance in
${}^6$Li. Here, though, alternative methods to access the bare
objects have been demonstrated. The resonance deals with
spin-1/2 atoms, the dressed molecules are triplets, and the atoms have a
net spin equal to zero. A measurement of the magnetic moment per
atom~\cite{JOC03} should~\cite{FAL04} therefore only see the bare
molecules. A laser spectroscopic probe that  couples only to the triplet
molecules likewise allows for a measurement of the bare-molecule
fraction~\cite{PAR05}.

An alternative angle helps with the interpretation
of the results. Let us initially take a single
zero-momentum bare molecule. The available state space is spanned by
the states with either one molecule at zero momentum and no atoms,
$b^\dagger_0|0\rangle$, or no molecule and pairs of
atoms
$\bk\!\!\uparrow$ and $-\bk\!\!\downarrow$,
$c^\dagger_{\bk\uparrow}c^\dagger_{-\bk\downarrow}|0\rangle$, where
$|0\rangle$ is the particle vacuum. We are at liberty to write the state
vector as
\begin{equation} |\psi(t)\rangle =
{1\over\sqrt{2}}\,\beta(t)b^\dagger_0|0\rangle + {1\over\sqrt N}
\sum_\bk C_\bk(t)
c^\dagger_{\bk\uparrow}c^\dagger_{-\bk\downarrow}|0\rangle\,.
\label{DRESSEDSTATE}
\end{equation}
Assuming self-consistently that the $C$ coefficients only depend on
energy,
$C_\bk(t)\equiv C(\epsilon;t)$, the Hamiltonian~\eq{HAM} gives
the time dependent Schr\"odinger equations for the coefficients $\beta$
and $C(\epsilon)$ that coincide with Eqs.~\eq{BEQEQ}
and~\eq{FINEQ}, except that in the counterpart of Eq.~\eq{BEQEQ}  we
have unity in lieu of the factor $[1-2P(\epsilon)]$. Were it not
for this factor that obviously reflects the Pauli exclusion principle,
our mean-field theory would represent a collection of independent
dressed molecules.

A stationary state of the form~\eq{DRESSEDSTATE}
is a molecule ``dressed'' with atom
pairs~\cite{MAC02,JAV04,MAC05,STO04,FAL04}. With the same renormalization
that was applied in Eq.~\eq{GAPEQ}, the dressed molecule is dissociated
for
$\delta>0$ and has one bound state for $\delta<0$. The latter is the
bound state that molecular spectroscopy~\cite{REG03,BAR05} detects in a
dilute gas. The effect of the renormalization is to put the
position of the Feshbach resonance in a dilute gas at the detuning
$\delta=0$.

The properties of the dressed molecule depend on the coupling function
$f(\epsilon)$, which may be obtained in principle from molecular
structure calculations. Setting $f(\epsilon)\equiv1$ corresponds to a
contact interaction for the conversion between atoms and molecules. In
reality, though, the interaction has a finite range.
Corresponding to a nonzero spatial range, there is
a range in energy; at high enough atom energies the coupling
between two bare atoms and a bare molecule must vanish. As
before~\cite{JAV04,MAC02}, we use here the model in which we simply cut
off the coupling at an energy
$\hbar M$ and write $
f(\epsilon) = \theta(M-\epsilon)\,,
\label{COUPLINGSHAPE}
$
where $\theta$ is the Heaviside unit step function. The binding energy
$2\hbar|\mu|$ of the dressed molecule for a negative detuning $-|\delta|$
may be solved from the transcendental equation
\begin{equation}
|\delta|-2|\mu|-{3\over4}
\sqrt{|\mu|}\sqrt{\bar\omega}\arctan\sqrt{M\over|\mu|} = 0\,,
\end{equation}
with
   $\bar\omega=\Omega^4/\epsilon_F^3$. While
$\Omega$ and $\epsilon_F$ depend on the density, in the
single-molecule parameter $\bar\omega$ the density dependence duly
cancels. Given the binding energy, the bare-molecule fraction $\beta^2$
in the dressed-molecule state vector is easily found.
A substantial molecular component, say,
$\beta^2>1/2$ emerges in the contact interaction case $M=\infty$ for
detunings $\delta<-(7\pi^2/512)\,\bar\omega~\sim -0.1\,\bar\omega$ and in
the case
$M\ll\bar\omega$ for detunings $\delta<-(3/4)\,\sqrt{M\bar\omega}$. Until
further notice all results are for the contact interaction, $M=\infty$.

{\em Main Features.}--In the limit of weak coupling,
$\Omega\ll\epsilon_F$, the relevant demarkation points for the
detuning $\delta$ are $0$ (Feshbach resonance) and
$2\epsilon_F$; 2 because it takes two atoms to make a molecule. For
$\delta<0$ the system is a condensate of
bare molecules, for $\delta>2\epsilon_F$ a
weak-coupling BCS superfluid. For $0<\delta<2\epsilon_F$ we have an
atom-molecule mixture in which all atoms from the Fermi sea above the
energy $\hbar\delta/2$ have been converted into bare molecules with this
same energy, so that $\beta^2=1-[\delta/(2\epsilon_F)]^{3/2}$. As has been
noted before~\cite{ZWI04,STO04,JAV04,WIL04,MAC05}, the Fermi sea of atoms
blocks the decay of the molecules even if the molecules are above the
dissociation threshold.

In the intermediate case with $\Omega\simeq\epsilon_F$ numerical results
show a continuous transformation from bare molecules to dressed
molecules, to a nondescript mixture of atoms and molecules, and
finally to a weak-coupling BCS as the detuning is varied from well below
the Feshbach resonance to well above it. Typical ${}^{40}$K
experiments~\cite{REG04} with $\Omega\simeq
4\epsilon_F$~\cite{JAV04,MAC05} belong to this category. In the present
approach the calculated fraction of bare molecules at the Feshbach
resonance is 6\%, and ca. 8\% for non-zero temperature \cite{MAC05,STO04},
enough to cast ambiguity on the observation~\cite{REG04} of fermionic
condensation.

For the 834~G Feshbach resonance in $^6$Li,
$\Omega\gg\epsilon_F$. The coupling is now so strong that dissociation of
above-threshold molecules is no longer Pauli blocked~\cite{MAC05}, which
may bring universal single-channel models into play. Since the term
universality comes in many shades, we frame the issues somewhat along the
lines of Refs.~\cite{DIE04,HO04,DEP04}, in the order of increasing
degree of universality. First, there are single-channel theories where
atoms and their interactions are the building blocks, and explicit
molecules are absent. The question is, do single-channel and two-channel
descriptions give the same results? Second, can the relevant atom-atom or
atom-molecule interactions be rolled into a single parameter, such as a
scattering length, so that the results of either theory or experiments
are independent of the microscopic physics of the interactions? Third,
given the interaction parameter, will the system reach a unitarity limit
that does not even depend on this parameter anymore as the interaction
strength increases?

Zero-temperature strong-coupling BCS theory is a single-channel
model characterized entirely by the scattering length, and gives results
that are independent of the scattering length when the scattering length
tends to infinity. In this strong form of
universality the BCS theory
agrees with experiments carried out close to the Feshbach
resonance~\cite{OHA02,BOU02,GUP02}, although it  may or may not be
quantitatively accurate~\cite{PER04,CAR03}. Moreover, as may be deduced
easily from the arguments of Ref.~\cite{DIE04}, in
the case
$\Omega\gg\epsilon_F$ and uniformly for all detunings $\delta>0$, our
two-channel mean-field theory and the standard BCS theory give the same
results. The product of Fermi wave number and scattering length,
$k_F a$, is the dimensionless scaling parameter of the BCS theory. The
translation into our model reads
$k_F a
\equiv-{3\pi\Omega^2/8\epsilon_F\delta} $.

On the other hand, on the side of large negative detunings and
in the limit of strong interaction,  the two-channel
theory scales with the dressed-molecule parameter
$\delta/\bar\omega$. This parameter cannot be expressed in
terms of $k_F a$ and experimental constants such as atom mass,
so that BCS and two-channel theories  are not equivalent.
As the normalization equation in the BCS theory is Eq.~\eq{NORMEQ}
sans the explicit $\beta^2$, an increasing fraction of bare molecules
signals  an increasing discrepancy between single- and
two-channel mean-field theories.

{\em Non-universal Experiments.}--Our two-channel theory covers all
parameter regions, and has different scaling properties on different
sides of the resonance. It is therefore of particular interest that the
Rice group has recently measured~\cite{PAR05} the bare-molecule fraction
of a Fermi gas over a wide range of magnetic fields across the
resonance. Comparisons between theory and experiment are
complicated by the fact that the experimental density varies
substantially between data points taken at different magnetic fields.
Density information can be recovered from a representation of the data in
terms of the parameter
$k_Fa$~\cite{PAR05}, but $k_Fa$ is not a scaling variable in our theory
and we cannot use it unambiguously  as the abscissa when plotting the
calculated results. For these reasons we use the
average of the Fermi energies deduced from the quoted values of $k_F a$
as our parametrization of the density. Since the interaction-strength
parameter scales weakly with density,
$\Omega/\epsilon_F\propto\rho^{-1/6}$, we still expect a qualitatively
valid comparison with experiments.

The Fermi energy here is given by $\epsilon_F = 7.46\times2\pi\, {\rm
kHz}$. The detuning in terms of magnetic field
$B$ is
$\delta=\Delta\mu(B-B_0)/\hbar$, where $\Delta\mu = 2\mu_B$ is the
difference in magnetic moments between a molecule and an atom pair, and
$B_0$ is the position of the Feshbach resonance. The standard relation
between atom-atom scattering length and magnetic field is $a =
a_b\left[1+{\Delta/ (B-B_0)}\right]$, where $a_b$ is the background
scattering length and $\Delta$ is the magnetic-field width of the
Feshbach resonance. For a comparison, we need to modify the
conversion of parameters between BCS and our approach to exclude the
background scattering length, so that
$k_F(a-a_b)=-{3\pi\Omega^2/8\delta\epsilon_F}$. The relation between
Fermi energy and wave number, of course, is
$\hbar\epsilon_F=\hbar^2 k_F^2/(2m)$. These observations, together with
the known resonance parameters~\cite{BAR05}, give the molecular parameter
$\bar\omega = 3.34\times 2\pi\,{\rm THz}$, enormously large
compared with any other frequency scale in the problem. Finally, the Rabi
frequency is given by
$\Omega/\epsilon_F=(\bar\omega/\epsilon_F)^{1/4} = 146$, which
indicates the strong-coupling regime.

\begin{figure}
\centering
\includegraphics[width=7.5cm]{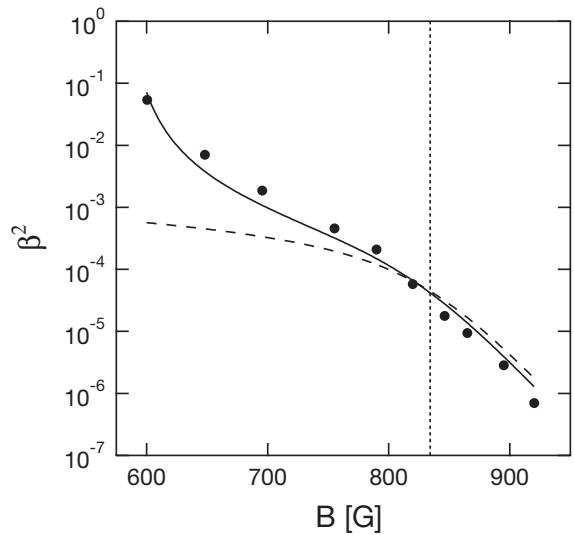}
\vspace{-0.3cm}
\caption{Bare molecule fraction in a ${}^6$Li gas as a function of
magnetic field from experiments~\protect\cite{PAR05} (circles), from our
theory with a contact interaction for atom-molecule conversion,
$M=\infty$,  (dashed line), and from the theory with energy cutoff
$M=254\times 2\pi\,{\rm kHz}$ corresponding to a range of the
interaction of about 1500~$a_0$ (solid line). The  dotted vertical line
marks the position of the Feshbach resonance.}
\label{EXPCOMP}
\vspace{-0.5cm}
\end{figure}

We plot in Fig.~\ref{EXPCOMP} the bare-molecule fraction $\beta^2$ as a
function of the magnetic field $B$ from the Rice data (circles) and from
our calculations assuming a contact interaction with $M=\infty$ (dashed
line). In the experiments the conversion into bare molecules extends a
few hundred Gauss below the resonance, while the contact-interaction
theory predicts a scale $\sim0.1\,\hbar\bar\omega/|\Delta\mu|\sim10\,{\rm
T}$ and completely misses a qualitative trend in the
experimental data. The remedy is to invoke a finite range for
atom-molecule conversion. We vary the parameter
$M$ until what is essentially the relative root-mean square error between
the experiments and the theory is minimized at $M=254\times 2\pi\,{\rm
kHz}$. The corresponding theory curve is plotted as the solid line. In
view of the qualitative assumptions about the Fermi energy in the
experiments and the simplistic model for the energy dependence of
atom-molecule coupling, the agreement between theory and experiment is
now excellent.

The magnetic-moment data~\cite{JOC03} on the bare-molecule
fraction covers a much smaller dynamical range than optical spectroscopy,
and only on the molecule side of the resonance. The
corresponding theory in Ref.~\cite{FAL04} is a
dressed-molecule argument for $\delta<0$ that explicitly
builds in the background scattering length
$a_b= -1405\,a_0$. Interestingly, when
converted into a length via
$M=\hbar/(m\ell^2)$, our fitted cutoff $M$ gives
$\ell=1500\,a_0$.

{\em Conclusions.}--A finite cutoff ($M$) indicates
that a contact interaction independent of the details of molecular
physics is not sufficient to reach even a qualitative agreement with the
experiments. The Rice data
manifestly contradicts universality. Besides, experimental detection of
a bare-molecule fraction $0<\beta^2<1$ within a presumably
molecular gas poses a severe challenge to any single-channel theory. A
single-channel model can be retrofitted to include molecules as bound
states of atom pairs, which would be the counterpart of our dressed
molecules. But what, then, are the separate bare molecules?

We have studied a simple two-channel mean-field
theory for a zero-temperature Fermi gas at a Feshbach resonance. The
model applies seamlessly across the resonance; no mixing or matching to
join the BCS and BEC regimes is needed, or allowed. Our results agree
with recent measurements of the fraction of bare molecules at various
magnetic fields~\cite{PAR05} over a dynamical range of five orders of
magnitude. Even though the
${}^6$Li  Fermi gas is strongly coupled, the experimental results are not
universal but reflect a finite range of atom-molecule coupling.

This work is supported in part by NSF (PHY-0354599) and NASA
(NAG3-2880). Randy Hulet generously provided both
data in a digital form, and valuable advise.

\end{document}